\documentclass[twocolumn,amsmath,amssymb]{revtex4}

\usepackage{graphicx}% Include figure files
\usepackage{dcolumn}% Align table columns on decimal point
\usepackage{bm}% bold math
\usepackage{epsf}
\usepackage{amsmath}
\usepackage{amssymb}
\usepackage{color}
\newcommand{\bra}[1]{\langle #1|}
\newcommand{\ket}[1]{|#1\rangle}
\newcommand{\braket}[2]{\left\langle #1|#2\right\rangle}
\newcommand{\tr}[1]{\mathrm{tr}\left\{#1\right\}}

\newcommand{\la}{\left\langle}
\newcommand{\ra}{\right\rangle}

\newcommand{\bla}{bla\\bla\\bla\\bla\\bla}
\newcommand{\more}{...more\\more\\more\\more\\more}

\newcommand{\PRA}{Phys. Rev. A }

\newcommand{\PRE}{Phys. Rev. E }
\newcommand{\PRX}{Phys. Rev. X }
\newcommand{\PRL}{Phys. Rev. Lett. }
\newcommand{\EPL}{EPL }

\newcommand{\NJP}{New. J. Phys. }
\newcommand{\NC}{Nature Comm. }

\begin{document}

\title{Shortcut-to-adiabaticity Otto engine: A twist to finite-time thermodynamics}
\author{Obinna Abah}\email[]{o.abah@qub.ac.uk} \author{Mauro Paternostro} \email[]{m.paternostro@qub.ac.uk}
\affiliation{Centre for Theoretical Atomic, Molecular and Optical Physics, Queen's University Belfast, Belfast BT7 1NN, United Kingdom}
%\affiliation{Department of Physics, Friedrich-Alexander-Universit\"at Erlangen-N\"urnberg, D-91058 Erlangen, Germany}
%\email{o.abah@qub.ac.uk}
%\author{Eric Lutz}

%\pacs{05.70.-a}

\begin{abstract}
We consider a finite-time Otto engine operating on a quantum harmonic oscillator and driven by shortcut-to-adiabaticity (STA) techniques to speed up its cycle. We study its efficiency and power when internal friction, time-averaged work, and work fluctuations are used as quantitative figures of merit, showing that time-averaged efficiency and power are useful cost functions for the characterization of the performance of the engine. We then use the minimum allowed time for validity of STA protocol  relation to establish a physically relevant bound to the efficiency at maximum power of the STA-driven cycle.
%\bla

\end{abstract}
\maketitle

\section{Introduction}
Heat engines were the fulcrum of the first industrial revolution and, remarkably, still play a major role in today's technological landscape, all the way down to the nanoscale. However, at such length-scale, quantum fluctuations and effects become relevant and their influences on the performance of thermodynamic devices should be treated {\it cum grano salis}~\cite{cam11}.

Recently, this realisation has led to the substantive development of a quantum-based framework for the thermodynamics of non-equilibrium processes and systems. The pathway towards the construction of a fully operative quantum engine has been paved by the demonstration of the first single-particle heat engine based on trapped-ion technology~\cite{ros16}. The perspectives for full-fledged quantum thermo-machines are promising. 

The efficiency of an engine, defined as the ratio of energy output to energy input, is maximum for adiabatic modified processes \cite{cal85,cen01,wu07}. Such maximum performance is however associated with vanishing power that hinders its any practical purposes~\cite{and84}. A major challenge is to design energy efficient thermal machines that deliver more output for the same input, without sacrificing power~\cite{aps08}. One of the ways to achieve this goal is to employ a shortcut-to-adiabaticity (STA) approach~\cite{tor13}, where a perfectly adiabatic process is mimicked by the use of a suitably arranged fast manipulation of the system, designed in a way to drive it towards the desired physical configuration and suppress any final-state excitation that might have been induced by the finite-time dynamics. Different STA techniques have been developed, from counterdiabatic driving (also known as transitionless quantum driving)~\cite{dem03,dem05,ber09,cam15} to local counterdiabatic driving~\cite{iba12,cam13,def14}, from methods based on the use of dynamical invariants~\cite{che10}, to the so-called fast-forward technique \cite{mas10,mas11}. The effectiveness of such approaches has been addressed in a significant number of experimental endeavours~\cite{sch10,sch11,bas12,bow12,wal12,zha13,an16,du16,mar16}, which have demonstrated the viability of STA-based approaches to quantum dynamical control. Remarkably, recent theoretical studies have shown that STA methods may be employed to enhance the performance of thermal heat engines~\cite{den13,tu14,cam14,jar16,cho16,aba17,aba18,li18}.

Despite such promising developments, the energetic cost of shortcut driving of STA techniques and their impact in the performance of quantum heat engines are not yet fully understood~\cite{kos17}. Recently, the cost of achieving the desired adiabatic state has been related to the time-energy uncertainty relation \cite{san16,zhe16,cou16,cui16,cam17} and  time-averaged excess of the work fluctuations~\cite{fun17,bra17}. Another approach that has been put forward to quantify the cost of STA approaches is based on the quantification of the energy invested to operate the controller and the system~\cite{tor17}. %The focus of studies so far has been to understand the efficiency and power output of the STA heat engines.

A specific scenario is of particular relevance in this context, namely the performance/efficiency of heat engines optimized to yield maximum power. As a significant case, it is worth considering the case of an Otto cycle, whose efficiency at maximum power in the adiabatic limit has been shown to corresponds to the so-called Curzon-Ahlborn efficiency~\cite{cur75,lef87,rez06,aba12}
\[
\eta_\text{CA} = 1 -\sqrt{\beta_C/\beta_H},
\]
where $\beta_C$ and $\beta_H$ %=(k_\textrm{B}T_i)^{-1}~(i=C,H)$ 
are the inverse temperature of a cold and a hot heat reservoir, respectively. This expression is not universal and depends on the assumption that the cycle time is constant. Thus, an interesting point is to understand the bounds imposed on the efficiency of engines such as the Otto one when operating at maximum power and driven by STA-based protocols.

Motivated by such observation and the somehow paradigmatic nature of the Otto cycle, in this paper we analyze the performance of a STA quantum Otto heat engine that uses a harmonic system as its working medium. We calculate the internal friction, time-averaged work, and work fluctuations to illustrate the energy cost of driving it through a STA technique. We further show the time-averaged efficiency and power of the engine are faithful and meaningful criteria to evaluate the performance of a STA-based engine. In addition, the bound on the efficiency at maximum power of STA Otto engine is derived considering the time imposed by the energy-time uncertainty relation on the system evolution during the STA protocol.

The remainder of this paper is organized as follows. In Sec.~\ref{QOC} we illustrate the non-equilibrium thermodynamics of a quantum Otto cycle, providing explicit formulae for work done, heat exchanged, and entropy produced during the relevant strokes of the cycle. Sec.~\ref{STAengine} is dedicated to the illustration of a STA-driven version of the cycle and the effect that the drive has on relevant thermodynamic quantities. In addition, we quantify the cost of such quantum control strategy  using a number of physically different figures of merit, including work friction and time-averaged work variance. In Sec.~\ref{bounds} we set physically rigorous bounds on the efficiency of the STA-driven cycle run at maximum power, showing the effectiveness of the quantum control strategy in achieving values of power and efficiency close to the adiabatic ones. Finally, in Sec.~\ref{conc} we draw our conclusions and set up the path to further investigations.

\section{Quantum Otto cycle}
\label{QOC}
\begin{figure}[!t]
\includegraphics[width=0.9 \linewidth]{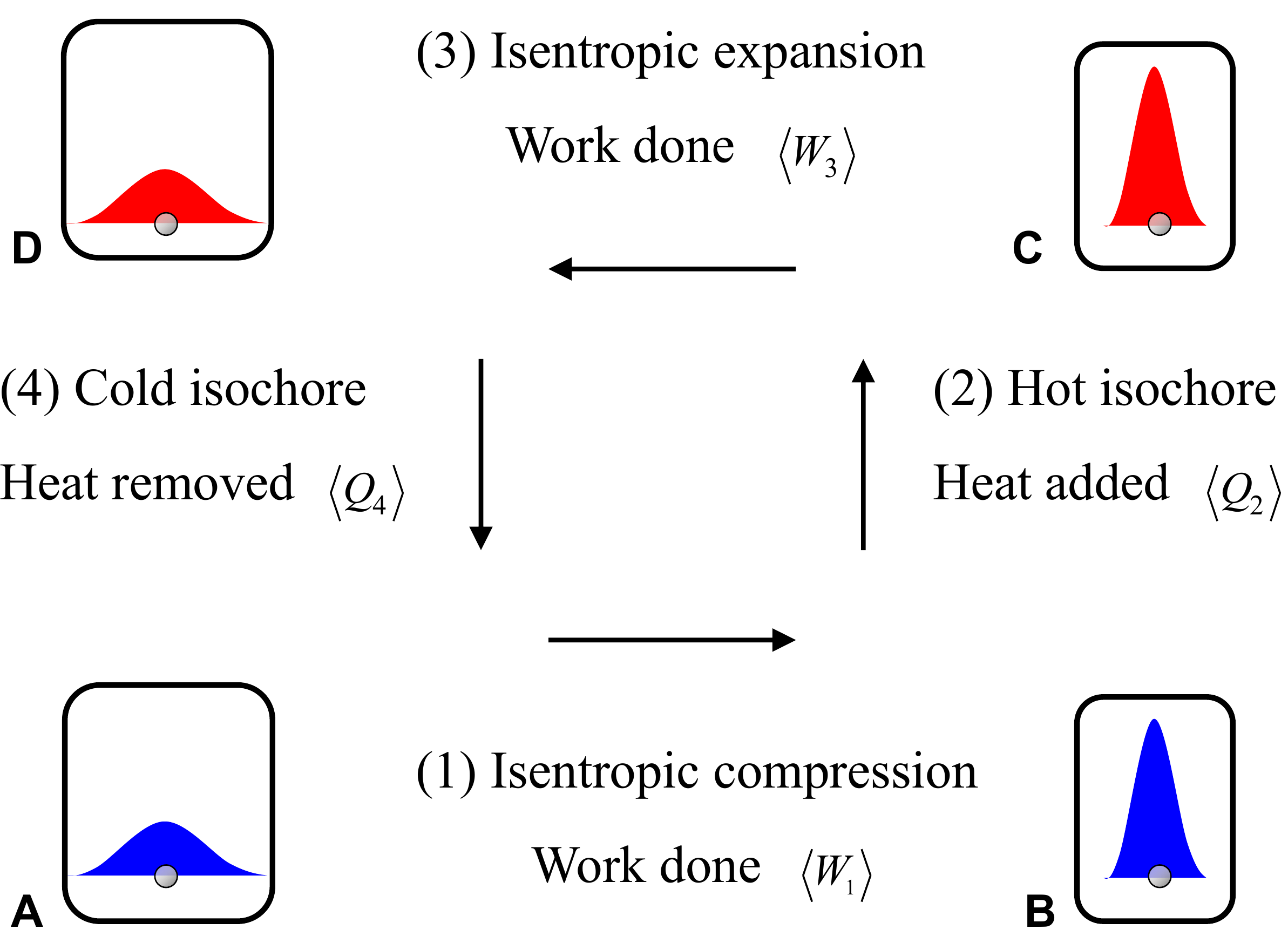}
\caption{Diagram of a quantum Otto cycle. The thermodynamic cycle consists of two isentropic  (compression and expansion steps 1 and 3) and two isochoric processes (heating and cooling steps 2 and 3). Here, $\langle W_j\rangle~(j=1,3)$ and $\langle Q_j\rangle~(j=2,4)$ stand for the average work done and heat exchanged during the relevant stroke of the cycle. The red- blue-colored areas represent the wave-function of the oscillator embodying the working medium of the cycle. }
\label{fig1}
\end{figure}
In a quantum Otto heat engine, the working medium undergoes a four-stroke cycle by being alternatively coupled to two baths at different temperatures. The Hamiltonian of the working medium $H (\lambda_t)$ depends on time-dependent work parameter $\lambda_t$ that determines the overall evolution of the medium.  
 As shown in Fig.~\ref{fig1}, the cycle is made of the following steps:
 \begin{itemize}
\item[(i)] An isentropic compression (branch $AB$ of the cycle), during which the working medium is isolated from the environment and its  work parameter $\lambda_t$ is increased  from $\lambda_1$ to $\lambda_2$ in a time $\tau_1$. As a result of this transformation, work $\la W_1\ra = \langle H\rangle_B - \langle H\rangle_A$ is performed on the medium. 
\item[(ii)] A hot isochore (branch $BC$ of the cycle), during which heat $\la Q_2\ra = \langle H\rangle_C - \langle H\rangle_B$ is transferred -- in a time $\tau_2$ -- from the hot bath at inverse temperature $\beta_2$ to the working medium. During such process, the working parameter takes the constant value $\lambda_2$.
\item[(iii)] An isentropic expansion (branch $CD$ of the cycle) where the work parameter is decreased -- in  a  time $\tau_3$ -- from value $\lambda_2$ to $\lambda_1$. During such transformation, an amount of work $\la W_3\ra = \langle H\rangle_D - \langle H\rangle_C$ is extracted from the medium. 
\item[(iv)] A cold isochore (branch $D A$ of the cycle) where heat $\la Q_4\ra = \langle H\rangle_A - \langle H\rangle_D$ is transferred -- in a time $\tau_4$ -- from the working medium to the cold bath at inverse temperature $\beta_1 > \beta_2$. The work parameter is again kept constant at value $\lambda_1$. 
\end{itemize}
The control parameters are the time-length of the different branches, the temperatures of the baths, and  the modulated frequency. We will assume, as it is  customary~\cite{lin03,rez06,qua07,aba12,kos17}, that the thermalization times $\tau_{2,4}$ are much shorter than the compression/expansion times $\tau_{1,3}$. The total cycle time is then  $\tau_\text{cycle}= \tau_1+ \tau_3=2 \tau$ for equal step duration.

For an engine,  the produced work  is negative, $\la W_1\ra +\la W_3\ra <0$, and the absorbed heat is positive, $\la Q_2\ra >0$. The two important quantities characterizing thermal machines are efficiency and power. The total change in entropy for one complete cycle reads
\begin{equation}
\Delta S_{tot} = -\beta_2 \la Q_2\ra - \beta_1 \la Q_4\ra \ge 0,
\label{2nd-law}
\end{equation}
where we used the fact that the entropy change during the isentropic processes $AB$ and $CD$ are zero.
From the first law of thermodynamics we have
\begin{equation}
-(\la W_1\ra + \la W_3\ra) = \la Q_2\ra + \la Q_4\ra.
\label{1st-law}
\end{equation}
In light of Eqs.~(\ref{2nd-law}) and (\ref{1st-law}), the efficiency of the cycle can be written as  
\begin{equation}
\eta_O = -\frac{\la W_1\ra + \la W_3\ra}{\la Q_2\ra} \le 1 - \frac{\beta_2}{\beta_1}  :=\eta_C.
\end{equation}
%where $\la W_{tot}\ra = \la W_1\ra + \la W_3\ra$.
That is, the efficiency is always less than Carnot efficiency $\eta_C=1-\beta_2/\beta_1$ and the equality holds only when $\beta_2/\beta_1 = \lambda_1/\lambda_2$. On the other hand, the power of the engine is given by the ratio of the work done to the time taken for one complete cycle, i.e.
\begin{equation}
P = -\frac{\la W_1\ra + \la W_3\ra}{\tau_{cycle}}.
\end{equation}

As the compression and expansion steps consist of both reversible and irreversible processes, the total work is $\la W_{tot}\ra = \la W_{ad}\ra + \la W_{irr}\ra$. The first term corresponds to the reversible (quasi-stationary) part of the transformation undergone by the working medium, while the second quantifies the inner friction of the process. The latter is connected to the quantum relative entropy between the density operator resulting from the non-equilibrium path and that associated with the hypothetical adiabatic one~\cite{def10a,cam14,pla14}. Explicitly, the work dissipated irreversibly along the non-adiabatic path reads 
\begin{equation}
\la W_{irr}\ra = S(\rho_t ||\rho_t^{ad})/\beta_t,
\end{equation}
where $S(\rho||\sigma) = \tr{\rho \ln \rho - \rho \ln \sigma}$ is the quantum relative entropy, $\rho_t$ is the instantaneous state at time $t$, and $\rho_t^{ad}$ is the corresponding adiabatic state.
%In general, the efficiency of the engine is bounded by $\eta \le \eta_{ad}$, where the equality holds in infinitely long time.
Further details on inner friction and irreversibility can be found in Refs.~\cite{ale15,fel03,cak16}.

%\subsection*{Linear engine: Harmonic oscillator}
Here we consider when the working medium is a time-dependent quantum harmonic oscillator. The corresponding Hamiltonian is of the standard  form, $H_0(\omega_t) = p^2/(2 m) + m\omega_t^2 x^2/2$, where $x$ and $p$ are the position and momentum operators of an oscillator of mass $m$. 

During the first and third strokes (compression and expansion), the  quantum oscillator is isolated and only work is performed by changing the frequency in time. As the dynamic is unitary, the Schr\"odinger equation for the parametric harmonic oscillator can be solved exactly for any given frequency modulation \cite{def08,def10}. The corresponding work values are  given by \cite{aba12}
\begin{equation}
\begin{aligned}
\la W_1\ra &= \frac{\hbar\omega_1}{2} (Q^\ast_1/x - 1) \coth\left(\frac{\hbar\beta_1\omega_1}{2}\right),\\
\la W_3\ra &= \frac{\hbar\omega_1}{2} (Q^\ast_3 - 1/x) \coth\left(\frac{\hbar\beta_2\omega_2}{2}\right),
\end{aligned}
\end{equation}
where we have introduced the frequency ratio $x={\omega_1}/{\omega_2}$ and the dimensionless adiabaticity parameter $Q^*_{i}$ $(i=1,3)$~\cite{hus53}. The latter is defined as the ratio of the instantaneous and corresponding adiabatic mean energy, and takes unit value for any adiabatic process~\cite{def10}. Its explicit expression for any frequency modulation $\omega_t$ may be found in Refs.~\cite{def08,def10}. On the other hand, the heat exchanged with the reservoirs during the thermalization step (the hot isochoric process) reads
 \begin{equation}
\la Q_2\ra = \frac{\hbar \omega_2}{2} \left[\coth\left(\frac{\hbar\beta_2\omega_2}{2}\right) - Q^\ast_1 \coth\left(\frac{\hbar\beta_1 \omega_1}{2}\right) \right].
\end{equation}

The exact engine efficiency and power read as follows \cite{aba12}
\begin{equation}
\eta_O = 1 - x \left(\frac{x\, Q^\ast_3\, \langle H\rangle_C - \langle H\rangle_A }{x\,\langle H\rangle_C - Q_1^\ast \langle H\rangle_A} \right),
\label{3}
\end{equation}
\begin{equation}
P =\left[ \langle H\rangle_A \left(1 - Q_1^\ast /x\right) + \left(1 - x\, Q_3^\ast\right) \langle H\rangle_C \right]/\tau_{cycle},
\label{4}
\end{equation}
where $\la H\ra_A =\hbar\omega_1 \coth(\hbar\beta_1\omega_1/2)/2$ and $\la H\ra_C = \hbar\omega_2 \coth(\hbar\beta_2\omega_2/2)/2$.
These expressions are exact and valid at arbitrary temperatures, frequencies and time length. In the limit of slow driving (i.e. when $\tau_{cycle}$ becomes very large and the cycle tends towards adiabaticity), during the isentropic processes $Q_i^\ast = 1$, and the engine efficiency reads $\eta_O^\text{AD} = 1 -x$, while the power vanishes.
However, it has been shown that the optimal performance corresponds to an adiabatic version of the first and third stroke of the engine cycle~\cite{rez06,aba12}.

\section{Shortcut-to-adiabaticity engine}
\label{STAengine}
The dynamics of the quantum Otto engine may be sped up with the help of STA techniques applied to the compression and expansion steps.
STA protocols suppress the unwanted nonadiabatic transitions, thereby reducing the associated production of entropy~\cite{fel00,rez06,aba12,aba16}. The effective Hamiltonian of the oscillator is  then of the form
\begin{equation}
H_\mathrm{eff}(t) = H_0(t) + H_\text{STA}^i(t),
\end{equation}
where $H_\text{STA}^i(t)$ is the STA driving Hamiltonian and $i =(1,3)$ indicates the respective compression/expansion step. The STA protocol satisfies boundary conditions  that  ensure $\langle H_\text{STA}^i(0)\rangle=\langle H_\text{STA}^i(\tau)\rangle=0$. These correspond to requesting
\begin{equation}
\label{5}
\begin{array}{lcr}
\omega(0) = \omega_i,& \dot{\omega}(0) =\ddot{\omega}(0) = 0,\\
\omega(\tau) =\omega_f,& \dot{\omega}(\tau) =\ddot{\omega}(\tau) = 0,
\end{array}
\end{equation}
where $\omega_{i,f}= \omega_{1,2}$ denote the respective initial  and final frequencies of  the compression/expansion steps. Eqs.~\eqref{5} are satisfied, for instance, by the following functional forms~\cite{che10,iba12,cam13,den13,dam14,def14}
\begin{equation}
\label{protocols}
\begin{aligned}
\omega^{(1)}(t)&= \omega_i + 10\delta s^3 - 15\delta s^4+ 6\delta s^5,\\
\omega^{(2)}(t) &= \omega_i + (3 s^2 - 2 s^3)\delta,\\
\omega^{(3)}(t) &= \omega_i \sqrt{[(a^2 + 1) - (a^2 - 1)\cos(\pi s)]/2},
\end{aligned}
\end{equation}
where $s = t/\tau$, $a = \omega_f/\omega_i$ and $\delta = \omega_f - \omega_i$. 

\begin{figure}[b!]
\includegraphics[width=0.95\linewidth]{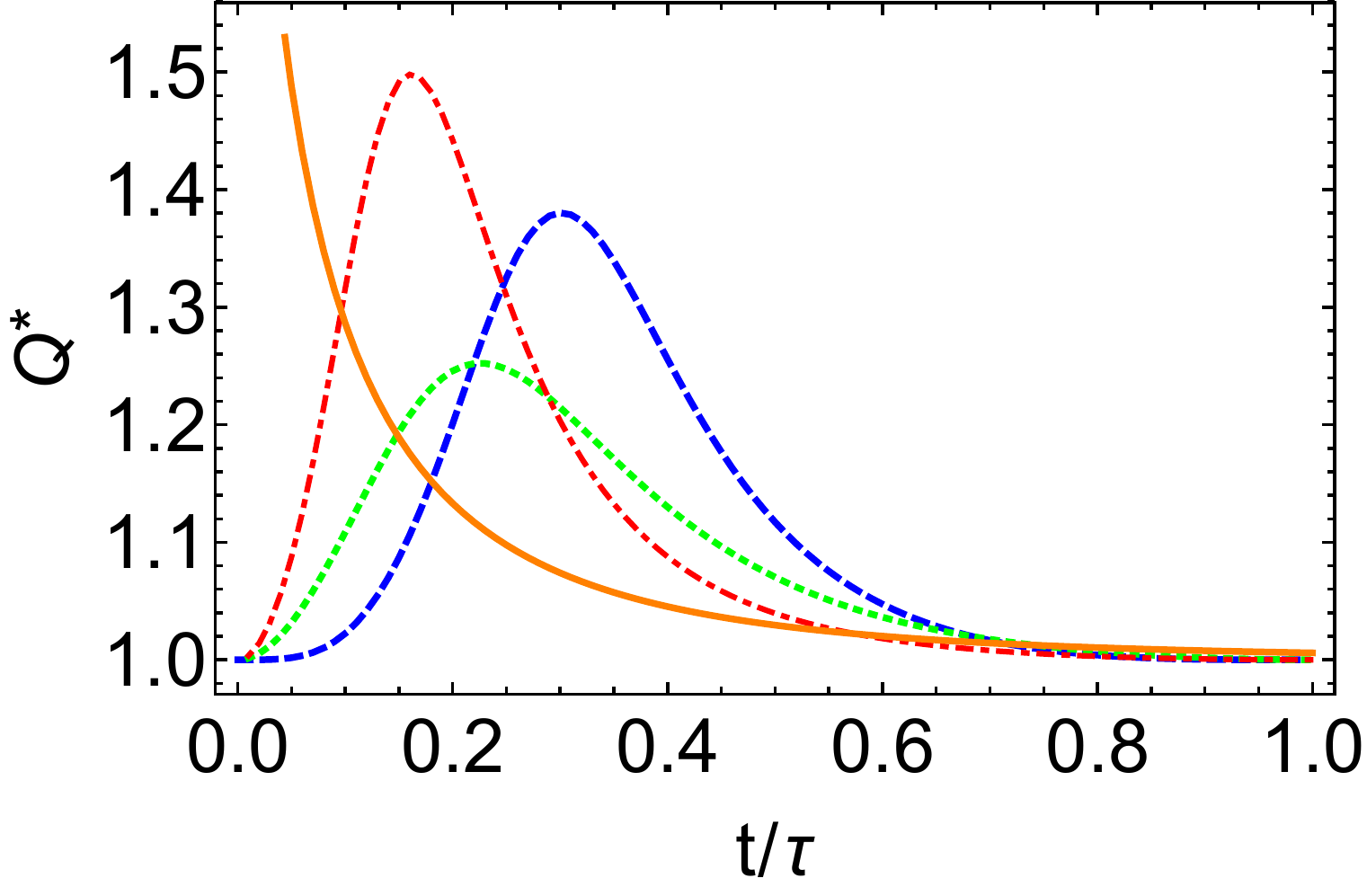}
\caption{The adiabaticity parameter for different time-dependent frequency protocol. The  blue dashed line is for $\omega^{(1)}(t)$, the green dotted line is for $\omega^{(2)}(t)$ and the red dotted-dashed line is for $\omega^{(3)}(t)$ in Eq.~\eqref{protocols}. The orange solid curve is the time-dependent frequency control $\omega_t = \omega_i + (\omega_f - \omega_i) t/\tau$, which does not satisfy the STA boundary conditions. The parameters used in all the curves shown in this plot are $\omega_i/\omega_f = 0.35$ and $\tau = 3 $ (in units of $1/\omega_f$) .}%We have taken units such that $\hbar=1$ in these calculations. 
\label{Qast}
\end{figure}
Let us consider when the compression/expansion steps are driven by a suitable STA approach. Here, we employed the counterdiabatic driving (transitionless quantum driving) whose goal  is to find a Hamiltonian $H_\text{CD}$ for which the adiabatic approximation to the original Hamiltonian $H_0$ is the exact solution of the time-dependent Schr\"odinger equation for $H_\text{CD}$. The explicit  form of $H_\text{CD}$ is~\cite{ber09,mug10}
\begin{eqnarray}
\label{9}
H_\text{CD}(t) &=& H_0(t) + i\hbar \sum_n(\ket{\partial_t n}\bra{n} - \braket{n}{\partial_t n} \ket{n}\bra{n}) \nonumber \\
&=& H_0(t) + H_\text{STA}^\text{CD}(t),
\end{eqnarray}
where $\ket{n} \equiv \ket{n(t)}$ denotes the $n^\text{th}$ eigenstate of the original Hamiltonian $H_0(t)$, $\ket{\partial_t n} \equiv \ket{\partial_t n(t)}$ and  $H_\text{STA}^\text{CD}(t)$ is the STA driving Hamiltonian. For a   time-dependent harmonic oscillator, the latter is given by~\cite{mug10,tor13}
\begin{equation}
H_\text{STA}^\text{CD}(t) = -\frac{\dot{\omega_t}}{4\, \omega_t}({x}{p}+ {p} {x}) = \frac{i \hbar \dot{\omega_t}}{4\, \omega_t}(a_t^2 - a_t^{\dagger 2}),
\end{equation}
where we used the notation shortcut $\omega_t\equiv\omega(t)$ and introduced the standard bosonic annihilation and creation operators $a\ket{n} = \sqrt{n}\ket{n-1}$ and $a^\dagger\ket{n} = \sqrt{n+1}\ket{n+1}$.
The Hamiltonian in Eq.~\eqref{9} is quadratic in ${x}$ and ${p}$, so it may be considered describing   a generalized harmonic oscillator with a non-local operator~\cite{ber85,mug10,che10}
\begin{equation}
H_\text{CD}(t) = \frac{p^2}{2 m}+ \frac{m\omega_t^2 x^2}{2}-\frac{\dot{\omega_t}}{4 \omega_t}(xp +px).
\label{10}
\end{equation}
The instantaneous eigenenergies of the Hamiltonian $H_\text{CD}(t)$ are given by~\cite{mis17,aba18}
\begin{equation}
E_n = \la H_{CD}(t)\ra = \hbar \omega_t Q_{CD}^\ast  \left(n+1/2\right), 
\end{equation}
where $Q_{CD}^\ast = 1/\sqrt{1 - (\dot{\omega}_t^2/4\omega^4_t)}$ is the STA protocol adiabaticity parameter and $n =1/(\exp(\beta_i \hbar \omega_i) -1)$ is the occupation quantum number. Here the label $i$ stands for the initial frequency and inverse temperature. The  mean average of the STA control at any time is~\cite{aba18}
\begin{equation}
\la H_{STA}^{CD}(t)\ra =  \frac{\omega_t}{\omega_i} (Q_{CD}^\ast - 1)\la H(0)\ra,
\end{equation}
where we used $\la H(0)\ra = \hbar\omega_i\coth(\beta\hbar\omega_i/2)/2$. % and the adiabatic parameter $Q^\ast\ge 1$ describes the evolution of the system which equals to unity at the initial and final states. 
The adiabaticity parameter during the compression process for the three protocols in Eq.~\eqref{protocols} are plotted in Fig.~\ref{Qast}.

\subsection{Energy cost of shorcut driving}
As STAs are designed so that the initial and final state corresponds to the adiabatic ones, one might naively think that their implementation is actually cost-free. Unfortunately, this is not the case, and the quantification of the energetic cost associated with STA driving has been investigated, recently, using various approaches.

First, based on an optimal control approach~\cite{siv12}, the time average of the difference between the mean work produced using the STA-modified and original Hamiltonian can be used as a quantifier. Quantitatively, we have %employed to quantify the cost as 
\begin{equation}
\la \delta W\ra_\tau = \frac{1}{\tau}\int_0^\tau \delta W dt,
\end{equation} 
where we have introduced the work difference $\delta W \equiv \la H_{STA}^{CD}(t)\ra =\la H_{CD}(t)\ra - \la H_0(t)\ra.$
Interestingly, this cost parameter corresponds to the time-averaged STA control Hamiltonian use when analysing the efficiency of the protocol as well as the range of validity of the control technique~\cite{aba17,cam14}. Explicitly, we have 
\begin{equation}
\la \delta W\ra_\tau = \la H_{STA}^{CD}(t)\ra_\tau.
\end{equation}

The second approach employs the time-average of the difference between the values of the variance of the work distribution corresponding to the effective Hamiltonian and the adiabatic counterpart of the original Hamiltonian~\cite{fun17}. It reads
\begin{equation}
\la \delta \Delta W\ra_\tau = \frac{1}{\tau}\int_0^\tau \delta(\Delta W) dt
\end{equation}
where $\delta(\Delta W)^2 = \la \Delta W_{CD}(t)^2\ra - \la \Delta W_{AD}(t)^2\ra$ and $\la \delta \Delta W\ra = \sqrt{\delta(\Delta W)^2}$. This cost functional is shown to relate with quantum speed limit of the evolution and allegedly gives a tighter bound~\cite{fun17}.

Finally, to understand the inner friction of the driving we consider the difference between the actual (nonadiabatic) work and the adiabatic one. That is
\begin{equation}
W_{fric} = \la W\ra_\text{NA} - \la W\ra_\text{AD} = \la H_t\ra - \la H_t\ra_\text{AD},
\end{equation}
where $\la W\ra_\text{NA}$ is the exact work calculated from the working medium dynamics at any given time.
\begin{figure}[!]
\includegraphics[width=0.95\linewidth]{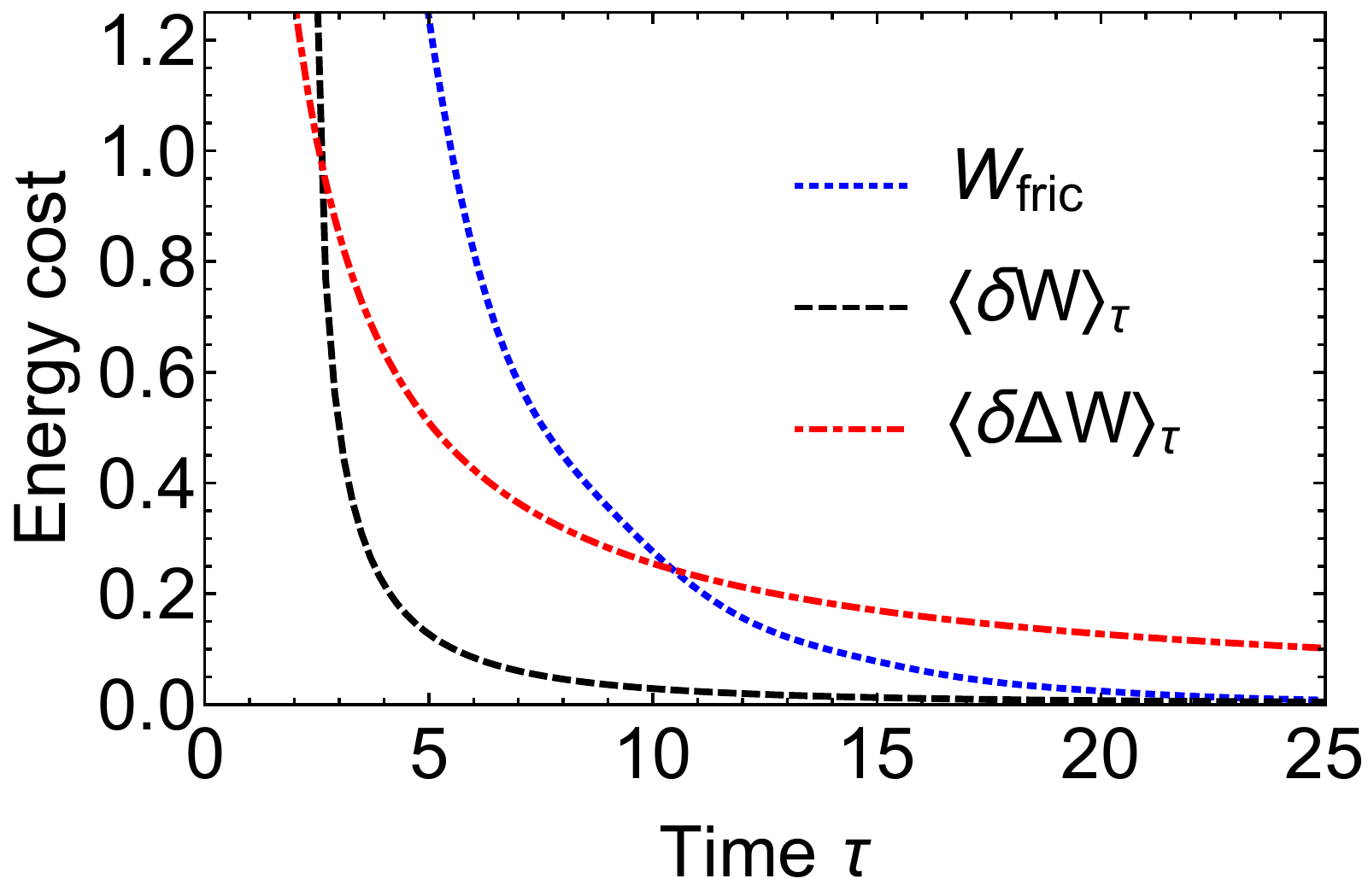}
\caption{The energetic cost (in unit of $\hbar \omega_f$) of STA driving as a function of  driving time $\tau$ (in units of $1/\omega_f$) for compression process.   We used the time-dependent frequency $\omega(t)= \omega_i + 10(\omega_f - \omega_i)s^3 - 15(\omega_f-\omega_i)s^4+ 6(\omega_f - \omega_i) s^5$ with the parameters  $\omega_i/\omega_f =0.35$, $\beta_2/ \beta_1 = 0.1$ and $\hbar =1$. }
\label{costcCD2}
\end{figure}

In Fig.~\ref{costcCD2} we show the time-averaged work difference and variance for a harmonic oscillator with a time-dependent frequency undergoing compression. We compare the result with the exact work friction at any given time, and the adiabatic work. In the example that follows, we show that the first definition, $\la \delta W\ra_\tau$, of STA protocol  vanishes at the point where the traditional friction is zero whereas the second definition is remains finite. This shows that the time-averaged work difference definition is the actual energetic cost of STA driven protocol and should be importance in thermodynamics analysis than the energy variance approach. Further analysis, of energy cost of the protocol $\omega_t = \omega_i + (\omega_f - \omega_i) t/\tau$ which does not satisfy the STA boundary conditions $\dot{\omega}_t = \ddot{\omega}_t = 0$ gives some finite term at $t =\tau$. This elucidates how evaluating the cost in the sense of a two-point energy measurement at the initial and final stages of the protocol does not give us meaningful results.     

\subsection{Engine performance}
Currently, there is no consensus on how to evaluate the cost of STA-based driving when considering the engine performance. This is mainly because of the boundary condition imposed by STA, [cf. Eq. \eqref{5}], which leads to vanishing of the adiabaticity parameter at the beginning and the final state of the driving protocol.  An approach adopted so far is taking the cost required for the control pulse as an additional energy input in efficiency~\cite{aba17,aba18} and which should be subtracted from the output power of the engine \cite{li18}.
Therefore, the efficiency of a  STA engine reads~\cite{aba17}
\begin{equation}
\eta_\text{STA} = \frac{\mathrm{energy \,output}}{\mathrm{energy\, input}} = -\frac{\sum_{j=0,1}\la W_{2j+1}\ra_\text{STA}}{\la Q_2\ra + \sum_{j=0,1} \langle H^{2j+1}_\mathrm{STA}\rangle_\tau}, \label{6}
\end{equation} 
where $\la W_i\ra_\text{STA} = \la H_\text{CD}(t)\ra - \la H(0)\ra$ is corresponding mean work of the STA protocol and  $\la H_\text{STA}^i\ra_\tau = (1/\tau) \int_0^\tau dt \la H_\text{STA}^i(t)\ra$ is the time-average of the mean STA driving term.
 Eq.~\eqref{6} takes the energetic cost of the STA driving  along the compression/expansion steps  into account. It
reduces to the adiabatic efficiency $\eta_\text{AD}$ in the absence of these two contributions as it assumes that $\la W_\text{STA}\ra = \la W_\text{AD}\ra$. 
 
The power produced by the STA-driven cycle is on the other hand given by the expression~\cite{li18}
\begin{equation}
P_\text{STA} = -\frac{1}{\tau_\text{cycle}}\sum_{j=0,1}\left(\la W_{2j+1}\ra_\text{STA}  - \langle H^{2j+1}_\mathrm{STA}\rangle_\tau\right). \label{7}
%P_\text{STA} = -\frac{\la W_{1}\ra_\text{STA} + \la W_{3}\ra_\text{STA} - \la H^1_\mathrm{STA}\ra_\tau - \la H^3_\mathrm{STA}\ra_\tau}{\tau_\text{cycle}}. \label{7}
\end{equation}
 At the initial and final time, the STA protocol ensures adiabatic work output, $\la W_{i}\ra_\text{STA} = \la W_{i}\ra_\text{AD}$ $(i=1,3)$. It has been shown that in a shorter cycle duration $\tau_\text{cycle}$, the superadiabatic power $P_\text{STA}$ is always greater than the nonadiabatic power $P_\text{NA}=-(\la W_{1}\ra + \la W_{3}\ra)/{\tau_\text{cycle}}$ \cite{aba17}. 

The STA technique is reminiscent of a periodic power signal which is zero at the beginning and the end of one complete cycle. The actual power of the cycle is thus customarily defined as the time-averaged one~\cite{hal13}. That is, despite the instantaneous performance of the STA engine seems the same as that of the adiabatic engine, their time-averaged performance is different. Thus, we define the time-averaged efficiency and power as 
\begin{equation}
\label{averaged}
\begin{aligned}
\la \eta_\text{STA}\ra_\tau &=-  \la\frac{-\la W_1\ra_\text{STA}+\la W_3\ra_\text{STA}}{\la Q_2\ra }\ra_\tau, \\
\la P_\text{STA}\ra_\tau &= -\frac{\la \la W_{1}\ra_\text{STA} + \la W_{3}\ra_\text{STA} \ra_\tau}{\tau_\text{cycle}}.
\end{aligned}
\end{equation}

\begin{figure*}[!]
{\bf (a)}\hskip5.5cm{\bf (b)}\hskip5.5cm{\bf (c)}\\
\includegraphics[width=0.64\columnwidth]{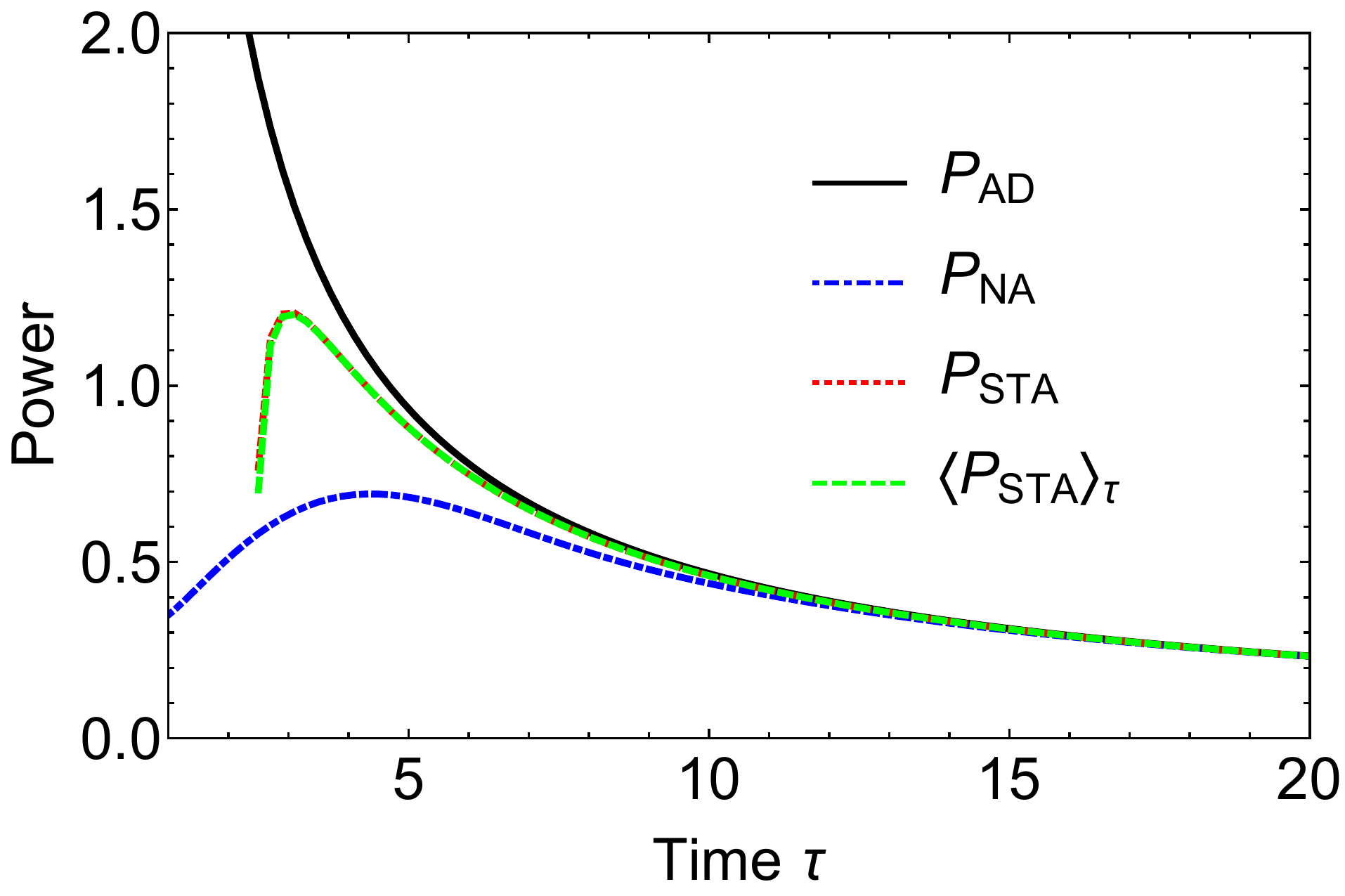}\hfil
\includegraphics[width=0.66\columnwidth]{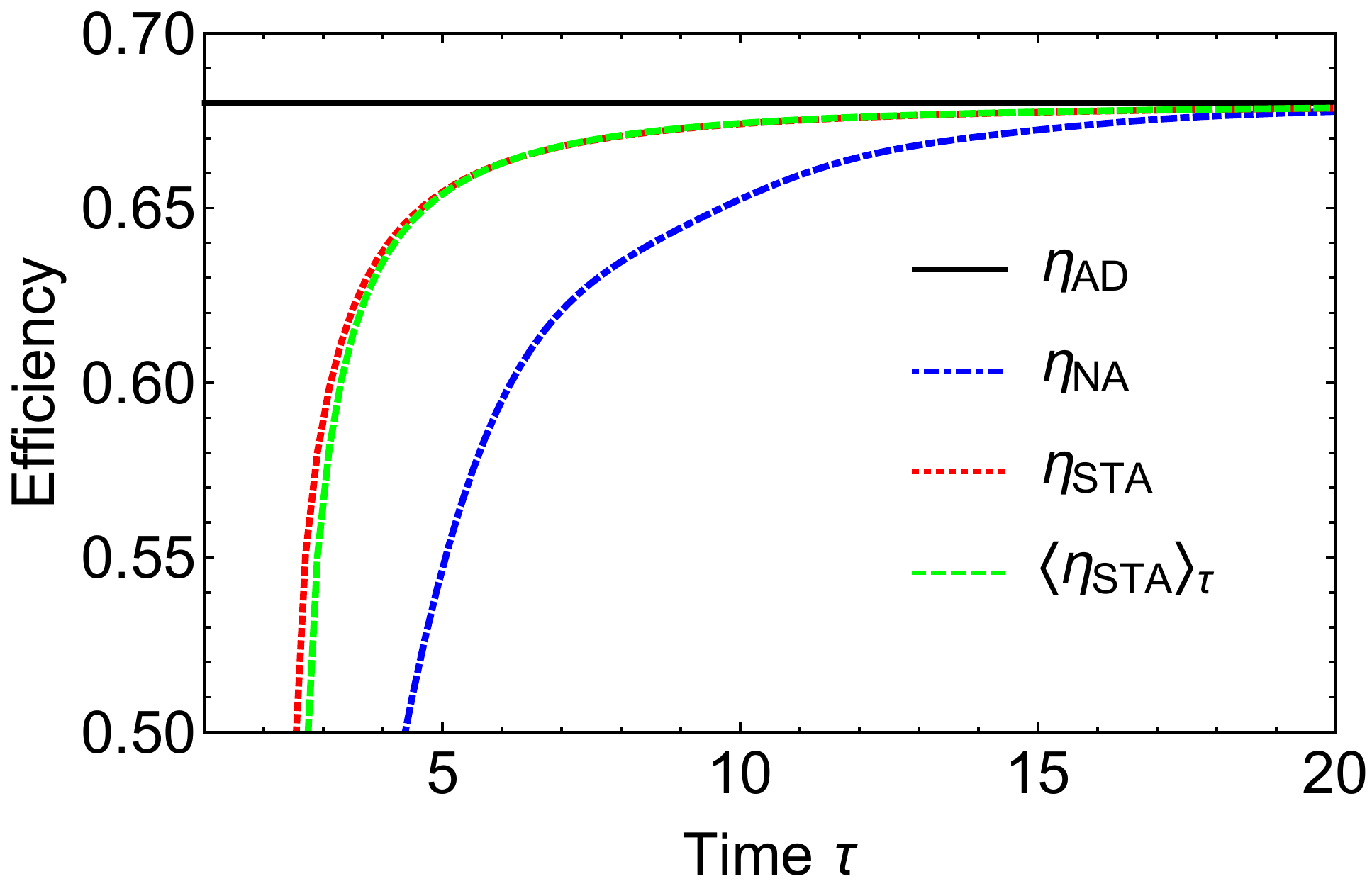}\hfil
\includegraphics[width=0.64\columnwidth]{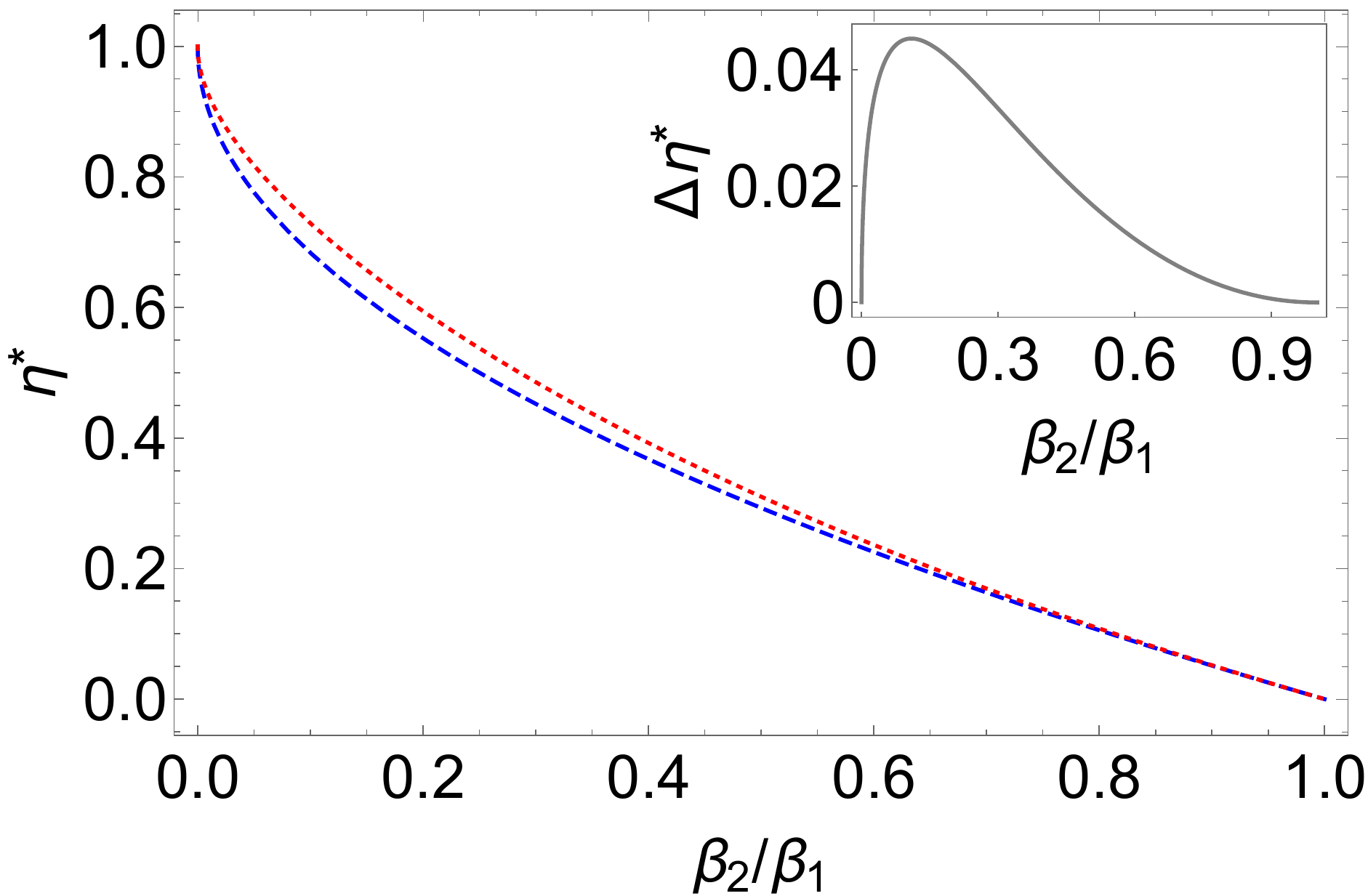}
\caption{Power [{\bf (a)}] and efficiency [{\bf (b)}] of the engine as a function of driving time $\tau $ (in  units of $1/\omega_f$). The black solid line is the adiabatic power [efficiency] when $Q^\ast_i = 1$ in Eqs.~\eqref{3} and \eqref{4}, while the blue dotted curve shows the nonadiabatic scenario without the STA Hamiltonian, cf. Eqs.~\eqref{3} and \eqref{4}. The red dotted curve shows the performance based on STA-driving (including the energetic cost of running the STA protocol), cf. Eqs.~\eqref{6} and \eqref{7}. Finally, the green dotted curve is the STA-driven performance based on time-averages [cf. Eqs.~\eqref{averaged}]. In our numerical evaluations, we have used the parameters $\omega_i/\omega_f =0.35, \beta_2/\beta_1 =0.1$, and $\hbar =1$. {\bf (c)} Efficiency at maximum power as a function of temperature ratio. We show the efficiency at maximum power of STA protocol $\eta^\ast_\text{STA}$ (red dotted) and compare it to the Curzon-Alhborn efficiency $\eta_\text{CA} =1 - \sqrt{\beta_2/\beta_1} $ (blue dashed). The inset shows their difference $\Delta \eta^\ast = \eta_\text{STA}^\ast - \eta_\text{CA}$. }
\label{power2}
\end{figure*}
Such quantities are presented in Fig.~\ref{power2}, where we observe that  the numerical evaluation of the time-averaged performance (efficiency and power) corresponds to the definitions taking the cost of STA into account. The little discrepancy/deviation in the efficiency plot is a result of taking the condition $Q^\ast = 1$ in the input heat $\la Q_2\ra$. Thus, taking the time-averaged of the exact efficiency and power for the Otto engine when considering the finite-time protocol gives the true performance at any given time.

\section{Bounds on performance: Efficiency at maximum power}
\label{bounds}

The efficiency at maximum power is a very informative figure of merit. Standard thermodynamic-cycle analysis is based on the concept of equilibrium, which implies quasi-static transformations and thus vanishingly small power outputs. Non-equilibrium thermodynamics, on the other hand, explicitly accounts for finite-time transformations that deliver a non-null power output, at the expense of the efficiency of the cycle. The idea is usually to incorporate the time dependence of heat transfer in the analysis of heat engine or to follow the engineers approach to calculate the so-called ``second-law or exergy efficiency" \cite{wu07}. However, for an Otto engine cycle, this approach leads to the assumption of constant (although finite) cycle time, thereby treating power as work~\cite{lef87,aba12,kos17} and leading to the efficiency at maximum power/work,  $\eta_{O}^\ast = 1 - \sqrt{\beta_2/\beta_1}\equiv \eta_{CA}$  in the high-temperature reservoirs limit. This is nothing else but the Curzon-Alhborn efficiency~\cite{cur75}.

When considering time-dependent transformation, additional constraints might have to be considered that could affect efficiency at optimal values of power. In our case, the STA-driven counterdiabatic protocol dynamics is valid only when $\omega_t^2(1- \dot{\omega}_t^2)/(4 \omega_t^4) > 0$, which has to be enforced in order to avoid the inversion of the harmonic trap~\cite{mis17,aba18}. 
For simplicity, we assume that the non-adiabatic excitations vanish at the end of the STA protocol, which correspond to $\la H_\mathrm{STA}^i\ra_\tau = 0$ and $Q^\ast =1$. We further assume that the time for isochoric processes are negligible. Then, the total time can be written as $\tau = \tau_1 + \tau_3 \simeq {1}/{\omega_1} + {1}/{\omega_2}$, where we used the condition that the final time and final frequency are inversely related, i.e $t_f = 1/\omega_f$. To compute efficiency at maximum power that is valid in both classical and semi-classical limits, we employ the power $P = -(\la W_1\ra_\mathrm{AD} + \la W_3\ra_\mathrm{AD})/\tau$.
%\begin{equation}
%P  =  \frac{\la H\ra_A(1-1/x ) + \la H\ra_C \left( 1-x \right)}{(1 + x)/{\omega_1}}.
%\end{equation}
By maximizing the resulting expression with respect to $x$ and fixing initial frequency $\omega_1$ and temperatures, we find the optimal ratio
\begin{equation}
x = \frac{\gamma_\beta + \sqrt{ 2\gamma_\beta (1+ \gamma_\beta)}}{2 + \gamma_\beta},
\end{equation}
where $\gamma_\beta = \la H\ra_A/\la H\ra_C$. The corresponding efficiency at maximum power reads
\begin{equation}
\eta_\text{STA}^\ast = 1 - \frac{\gamma_\beta +    \sqrt{ 4 \gamma_\beta (1 + \gamma_\beta)}}{2 + \gamma_\beta}.
\end{equation} 
This expression is valid for any cold-reservoir temperature and for both reservoirs in the high-temperature limit. 
In Fig.~\ref{power2} {\bf (c)}, we show $\eta^*_\text{STA}$ and compare it to the Curzon-Ahlborn efficiency that will result for fixed time~\cite{rez06,aba12,kos17} as the temperature ratio $\beta_2/\beta_1$ varies. % their difference is given by $\Delta\eta^\ast = \eta_{STA}^\ast - \eta_{CA}$.
Clearly, $\eta^*_\text{STA}$ is very close to such bound, showing the effectiveness of the constrained STA approach to deliver high-efficiency cycles outputting maximum power. The small discrepancy between the two quantities is due to the fact that the Curzon-Alhborn efficiency of the Otto engine  was derived based solely on work output \cite{lef87,rez06,aba12}. Thus, including the finite-time duration of the work (isentropic) branch in the performance at maximum power analysis of the engine gives the actual bound.    This contributes  to the discussion of non-universality of the efficiency at maximum power of irriversible heat engines.

\section{Conclusions}
\label{conc}
We have performed a detailed study of a STA Otto  heat engine showing that the nonadiabatic transition between the initial and final state of the engine depends on the chosen driving protocol. We have calculated internal friction, time-averaged work, and work fluctuations of the heat engine with the aim of  understanding the energetic cost of an STA driving. We have shown that the STA-based heat engine performance are better computed by the time-averaged efficiency and power. Furthermore, we have derived the  bound on the efficiency at maximum power of the STA-based heat engine based on the  minimum allowed time for validity of STA protocol. Our study provides a suitable and rigorous route to analyze the performance of STA Otto engine for any kind of driving protocol, which will hopefully stimulate  the development of STA for thermodynamics applications.

\section*{Acknowledgement}
OA thanks KITP, University of California Santa Babara for hospitality  supported in part by the National Science Foundation under Grant No. NSF PHY-1748958. 
The authors acknowledge the support by the Royal Society (grant numbers NF160966 and NI160057), the Royal Commission for the Exhibition of 1851, the SFI-DfE Investigator Programme grant (grant 15/IA/2864), and the H2020 Collaborative Project TEQ (Grant Agreement 766900).

\end{document}